\documentclass[useAMS,usenatbib]{mn2e}

\def\aap{AA}

\def\apjl{ApJL}
\def\mnras{MNRAS}
\def\apj{ApJ}
\def\aj{AJ}

\def\RE{{R_{\rm E}}}
\def\Rap{{R_{\rm ap}}}
\def\Reff{{R_{\rm e}}}
\def\Sigmacr{{\Sigma_{\rm cr}}}

\usepackage{graphicx}
\usepackage{float}
\usepackage{amssymb}
\usepackage{amsfonts}
\usepackage{amsmath} 
\usepackage{color}

\title[Lensing and Dynamics in Two Simple Steps]
      {Lensing and Dynamics in Two Simple Steps}

\author[Agnello, Auger \& Evans]{A. Agnello$^{1}$\thanks{E-mail:
    aagnello,mauger,nwe@ast.cam.ac.uk}, M.W. Auger$^{1}$,
  N. W. Evans$^{1}$\\ $^{1}$Institute of Astronomy, University of
  Cambridge, Madingley Road, Cambridge CB3 0HA, UK}

\begin{document}
\date{Accepted . Received }

\pagerange{\pageref{firstpage}--\pageref{lastpage}} 

\maketitle
\label{firstpage}

\begin{abstract}
We present a ready-to-use method to constrain the density distribution
in early-type galaxy lenses. Assuming a power-law density profile,
then joint use of the virial theorem and the lens equation yields
simple formulae for the power-law index (or logarithmic density
gradient). Any dependence on orbital anisotropy can be tightly
constrained or even erased completely. Our results rely just on
surface brightnesses and line-of-sight kinematics, making deprojection
unnecessary. We revisit three systems that have already been examined
in the literature (the Cosmic Horseshoe, the Jackpot and B1608+656)
and provide our estimates.  Finally, we show that the method yields a
good approximation for the density profile even when the true profile
is a broken power-law, albeit with a mild bias towards isothermality.
%using broken power-law density
%profiles, we show that the method still yields good results, albeit
%with a mild bias towards isothermality.
\end{abstract}

\begin{keywords}
Gravitational lensing: strong - Galaxies: elliptical and lenticular,
cD - Galaxies: kinematics and dynamics - dark matter
\end{keywords}

\section{Introduction}

When a galaxy acts as a gravitational lens, its mass can be probed
independently via analyses of lensing properties and stellar
kinematics. This combination has proved a powerful way of breaking
both the mass-sheet degeneracy in lensing and the mass-anisotropy
degeneracy in dynamics. It was first tackled by combining the lens
equation with the Jeans equations \citep[cf][]{Tr04}, usually with the
assumption of a constant velocity anisotropy profile.  Of course,
Jeans analyses give no guarantee that the chosen models correspond to
physically consistent distribution functions
\citep[cf][]{An09,Ci10}. Latterly, sophisticated methods to build
distribution functions of lensing galaxies using the Schwarzschild
method have been developed~\citep[e.g.][]{Ba07}. In both cases,
though, observational data are often not enough to constrain all of
the parameters used.

Here, we provide a simple method for combining lensing and
dynamics. Our starting point is the virial theorem, which has the
significant advantage that no assumption as to the anisotropy profile
is ever needed. As a concrete case, we examine the scale-free density
profiles $\rho\propto r^{-\gamma}$. Given an accurate photometric
profile and an estimate of the Einstein radius from strong lensing,
our method straightforwardly yields the value of the logarithmic
gradient of the density $\gamma$. It does not require any photometric
deprojection and it is robust under changes in orbital structure of
the lens galaxy. By considering three benchmark lenses -- the Cosmic
Horseshoe, the Jackpot and B1608+656 -- we demonstrate that our method
performs very creditably against more sophisticated algorithms. Given
its simplicity, we believe it provides the natural first step in
combining lensing and dynamics.

\section{The Virial Theorem and the Lens Equation}

One global constraint for galaxy lenses is given by the projected mass
within the Einstein radius. Another is given by the virial theorem,
which links the average kinematics of stars to the total mass density
and luminosity. In other words, lensing probes projected masses, while
photometry and line-of-sight motions together trace the
three-dimensional potential, so a combined analysis allows us to infer
properties of the total mass profile. Here, we will quantify this
framework by exploiting only the direct observables: surface
brightnesses and projected kinematics.

\subsection{A Global Constraint}

The basic observables are the Einstein ring radius $\RE$, the
surface brightness profile of the lens galaxy $\Sigma(R)$,
together with the mean $\overline{v}_{\rm los}(R)$ and mean-square
velocities $\overline{v^{2}}_{\rm los}(R)$ of stars along the line of
sight at projected radius $R$.

Let us start with the lens equation.  For a spherical lens, this gives
the projected mass within the Einstein ring radius as
\begin{equation}
M_{\rm p}(\RE)=\pi\Sigmacr\RE^{2},
\end{equation}
where $\Sigmacr$ is the critical surface density~\citep[see
  e.g.,][]{Sc92} For a power-law lens with an assumed total (i.e.,
luminous and dark matter) density $\rho = \rho_0 (r/r_0)^{-\gamma}$,
the projected mass within the Einstein ring $M_{\rm p}(\RE)$ reduces
to a standard integral which can be done analytically.  This yields:
\begin{equation}
\rho_{0}r_{0}^{\gamma}=(1-\kappa)\Sigmacr \RE^{\gamma-1}\frac{(3-\gamma)\Gamma(\gamma/2)}{2\sqrt{\pi}\Gamma((\gamma-1)/2)}\ ,
\label{eq:pllens}
\end{equation}
where we have allowed for the presence of an external mean convergence
$\kappa$ inside the Einstein radius \citep[e.g.,][]{suy10}. The
external shear, even if important in modelling the lens, does not
enter the mass normalization.

The second ingredient is the virial theorem.  For a spherical system,
the virial theorem can be formulated in terms of the surface
brightness as~\citep[][]{Ag12}
\begin{equation}
\langle\overline{v^{2}}_{\rm los}\rangle_{L}=\frac{16\pi
  G}{3L_{\rm T}}\int_{0}^{\infty}\Sigma(R)R\int_{0}^{R}\frac{\rho (r)r^{2}}{\sqrt{R^{2}-r^{2}}}\mathrm{d}r\mathrm{d}R
\label{eq:pvt}
\end{equation}
Here, $L_{\rm T}$ is the total luminosity, and we have introduced the
notation
\begin{equation}
\langle\mathcal{A}\rangle_{L}\equiv
{2\pi \over L_{\rm T}} \int_{0}^{\infty}{\mathcal{A}\Sigma(R)R}\mathrm{d}R .
\end{equation}
to denote luminosity averages.

For power-law density profiles, eqs~(\ref{eq:pllens}) and
(\ref{eq:pvt}) give us immediately
\begin{equation}
\frac{\langle\overline{v^{2}}_{\rm los}\rangle_{L}}{c^{2}}\frac{D_{l}D_{ls}}{D_{s}R_{E}}=\frac{1-\kappa}{3\pi}\mathcal{F}(\gamma)\frac{\langle R^{2-\gamma}\rangle_{L}}{R_{E}^{2-\gamma}}\ .
\label{eq:virialens}
\end{equation}
Here $D_{\rm s},$ $D_{\rm l}$ and $D_{\rm ls}$ are the
angular-diameter distances to the source, to the lens and from lens to
source respectively. The function $\mathcal{F}$ is analytic, namely
\begin{equation}
\mathcal{F}(x)=\frac{\Gamma(x/2)\Gamma((5-x)/2)}{\Gamma(2-x/2)\Gamma((x-1)/2)}\ .
\end{equation}

It is worth pausing to examine eq~(\ref{eq:virialens}), which is the
first main result of this {\it Letter}. Every quantity is known, or
can be computed directly from the observations, with the sole
exception of $\gamma$. Hence, the power-law index $\gamma$
depends just on global, measurable quantities. There is no dependence
whatsoever on the orbital anisotropy.

The magnitude of the external convergence $\kappa$ is affected by the
mass-sheet degeneracy \citep{fal85}. Following the results of
\citet{suy10}, we adopt a prior distribution for $\kappa$ with
zero mean and $\approx0.05$ dispersion. This translates into a
$\approx5\%$ uncertainty in the mass-normalization, or $\approx2.5\%$
in the velocity dispersion. The latter is small compared to the
typical $\approx 15\%$ statistical uncertainties of present-day
measurements.

%\textbf{Here we have chosen to remain within the common approximation
%  of spherical symmetry for both the luminous and dark
%  components. Although extensions can be derived for elliptical mass
%  distributions, the effect of flattening on the density exponent is
%  generally secondary. This can be verified e.g. by comparing the
%  results of spherical Jeans analyses in \citet{Au10} with those of
%  the axisymmetric models of \citet{Ba11}.  }

\begin{figure}
	\centering
        \includegraphics[width=0.45\textwidth]{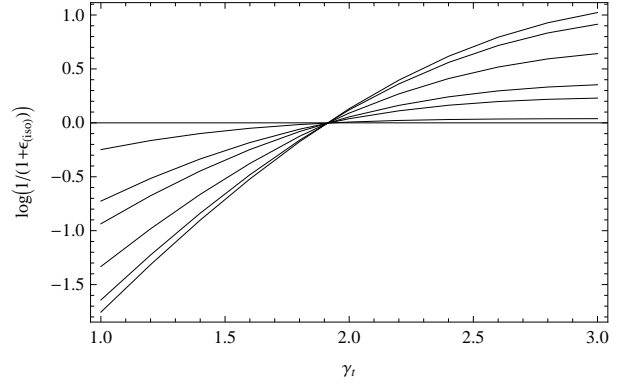}
\caption{\small{Logarithm of the aperture correction $1/(1+\epsilon)$
    as a function of $\gamma$. The lines are plotted for a de
    Vaucouleurs surface photometry and $\Rap/\Reff= 0.5, 0.6, 1, 2, 3,
    10.$ Smaller aperture radii correspond to steeper aperture
    corrections.}}
\label{fig:epsiso}
\end{figure}

\subsection{Aperture Corrections}\label{sect:apcorr}

In practice, the kinematics are measured inside a finite aperture
radius $\Rap$ and blurred by a point-spread function (PSF).  Hence, we
re-state the results in terms of the aperture-averaged line of sight
motions, introducing an aperture-correction:
\begin{equation}
\langle\overline{v^{2}}_{\rm los}\rangle_{L}=\langle\overline{v^{2}}_{\rm los}\rangle_{L(<\Rap)}(1+\epsilon)\ .
\label{eq:apcor}
\end{equation}
The function $\epsilon$ characterises how the aperture-average is
related to the luminosity-average over the whole system. If the
kinematics are measured well beyond the effective radius, or if the
latter is small compared to the fiber aperture, then the correction is
small and $\epsilon \approx 0$. For lenses in the SLACS sample, $\Rap$
is $\approx3/5$ of the de Vaucouleurs effective radius $\Reff$
\citep[cf][]{Au10}. Hence we need to inquire whether averaging within
the aperture radius (rather than to infinity) introduces an effective
dependence on orbital anisotropy. Let us assume, as a starting point,
that the system is isotropic within the effective radius, as suggested
by dynamical analyses of nearby elliptical galaxies
\citep{Sa93,Ge01}. We regard this as an informed prior on orbital
structure and examine in a second step the consequences of
anisotropy. We integrate the isotropic Jeans equations and
reconstruct the kinematic profile to quantify how much of it is left
out of the aperture-average. For a $\delta$-function PSF, the result
can be stated explicitly in terms of the surface brightness, namely
%:
\begin{equation}
1+\epsilon =\frac{L_{<\Rap}}{L_T}\left(
\frac{I(0,\gamma)}{I(0,\gamma)-I(\Rap,\gamma)}\right)\ ,
\label{eq:btbeta}
\end{equation}
were $L_{<\Rap}$ is the luminosity from projected radii less than
the aperture radius $\Rap$. We have also defined
\begin{equation}
I(y,\gamma)=\int_{y}^{\infty}\Sigma(R)R^{3-\gamma}g(y/R,\gamma)\mathrm{d}R\ ,
\end{equation}
where the kernel $g(y,\gamma)$ is given by
\begin{eqnarray}
\nonumber g(x,\gamma)=(4-\gamma)\int_{x}^{1}\sqrt{\frac{u^{2}-x^{2}}{1-u^{2}}}u^{3-\gamma}\mathrm{d}u\\
+(\gamma-1)x^{2}\int_{x}^{1}\sqrt{\frac{u^{2}-x^{2}}{1-u^{2}}}u^{1-\gamma}\mathrm{d}u\ .
\end{eqnarray}

Eq~(\ref{eq:btbeta}) is the second main result of this {\it
  Letter}. Here, the aperture correction is expressed as a double
quadrature that depends on the power-law density exponent
$\gamma$. Fig.~\ref{fig:epsiso} shows the aperture correction
$1/(1+\epsilon)$ for a de Vaucouleurs profile with different values of
$\Rap/\Reff$. For lenses in the SLACS sample, $\gamma \approx 2$, so
that the correction is generally steeper in $\gamma$ than the
right-hand side of eq~(\ref{eq:virialens}) and only weakly dependent
on the details of photometry. This behaviour is understandable, since
the additional requirement of pressure isotropy reduces the freedom in
$\gamma$ to yield physically consistent solutions.

The aperture corrections also can be evaluated for anisotropic velocity
dispersions, such as
\begin{equation}
\beta(r)= 
1 - {\overline{v^{2}}_{\theta} \over \overline{v^{2}}_r}
= {\beta_{0}\frac{r^{2}}{r^{2}+r_{a}^{2}}}.
\end{equation}
The systematic error on $\gamma$ given by neglecting anisotropy is
limited in modulus to $\lesssim 5\%$ for the SLACS sample, when
$\beta(\Reff)$ varies between -0.5 and 0.5 and a de Vaucouleurs
luminous profile is used. The correction to $\epsilon$ is always
negative for $\beta_{0}>0$ (radial anisotropy) and positive for
$\beta_{0}<0$ (tangential anisotropy). This is an extension of the
result found in \cite{Ko09}, where Hernquist or Jaffe luminous
densities with uniform anisotropy were used to quantify the effect. We
conclude that at least as far as aperture corrections are concerned,
velocity anisotropy is a secondary effect.

\subsection{Algorithm}

Eqs (\ref{eq:virialens}) and (\ref{eq:btbeta}) provide a prediction
for the root-mean-square velocities along the line of
sight, for which it is convenient to introduce the shorthand
\begin{equation}
v(\gamma) \doteq \sqrt{\langle\overline{v^{2}}_{\rm los}\rangle_L}.
\end{equation}
On the other hand, observations give us a value and uncertainties for
the same observable, say $\sigma\pm\delta\sigma$.

\begin{figure}
       \centering
\includegraphics[width=0.4\textwidth]{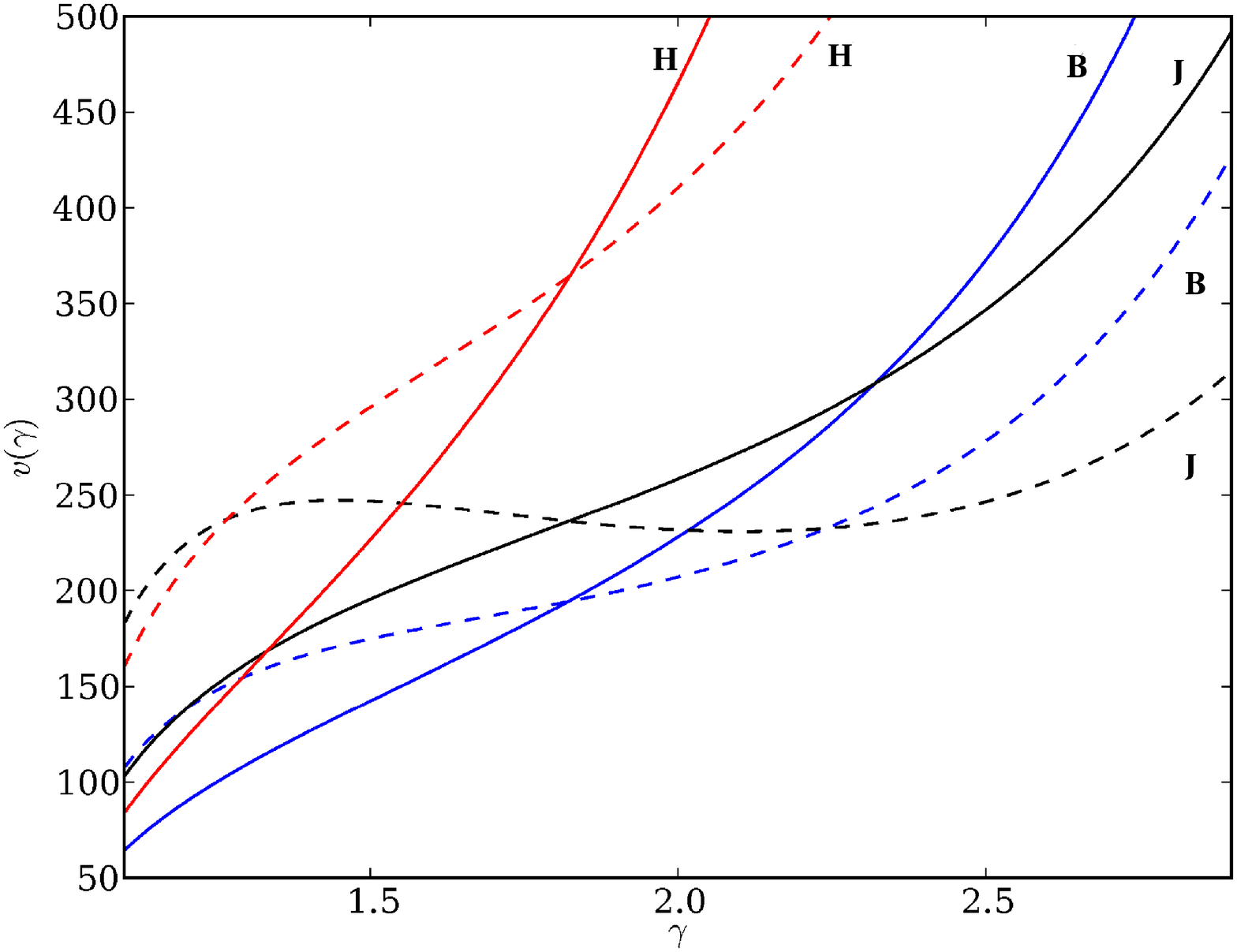}
\caption{\small{The relationship between the model power-law index
    $\gamma$ and the model velocity dispersion $v$ without aperture
    corrections (dashed) and including aperture corrections (solid)
    for the 3 example lenses discussed in Section 3 (red, labelled H):
    Horseshoe; black, labelled J: Jackpot; blue, labelled B:
    B1608). In addition to large quantitative changes in the predicted
    dispersions, the aperture corrections also yield significant
    qualitative changes; in all cases the relationship gets steeper
    when the aperture correction is applied.  This leads to more
    precise inference, particularly in the case of the Jackpot, where
    the flatness of the uncorrected relation implies little
    information on the slope given an observed dispersion, and the
    non-monotonic nature of $v(\gamma)$ also leads to a (biased)
    bi-modal inference.}}
\label{fig:gammavssigma}
\end{figure}

With larger fiber apertures, or extended long-slit measurements, there
is no need to make any aperture corrections.  So, for each value of
$\gamma$, eq~(\ref{eq:virialens}) yields the virial prediction and the
likelihood on the density exponent is simply
\begin{equation}
p(\gamma)\propto\exp- {[v(\gamma)-\sigma]^{2} \over 2\delta\sigma^{2}}.
\end{equation}
If the kinematics are not measured over a sufficiently wide radial
range, then both sides of eq~(\ref{eq:virialens}) are divided by the
aperture correction $1+\epsilon$, as defined in eq~(\ref{eq:btbeta}),
to get the new prediction for the line of sight velocity dispersions
and the new likelihood for $\gamma$. This step is critical when the
aperture is a small fraction of the effective radius, as $v(\gamma)$
may be very flat or even non-monotonic (Fig.~\ref{fig:gammavssigma});
in this case, if the aperture correction is ignored then the inference
on $\gamma$ may be biased and imprecise.

As illustrated, the combination of lensing and dynamics has been
reduced to the routine procedure of evaluating a single non-linear
function. When the PSF has a non-negligible width, the
aperture-averages are more representative of the luminosity
averages. In that case, the aperture corrections are smaller than
predicted by eq~(\ref{eq:btbeta}), which assumes a $\delta$-function
PSF. The density exponent must then lie between the uncorrected and
aperture-corrected estimates.

\section{Examples}

We first discuss three particular lens systems, comparing the
performance of our method\footnote{Code to implement our algorithm is
  available at \tt{http://www.ast.cam.ac.uk/\textasciitilde
    aagnello/virialLensing/}} with results in the literature. The
first quoted error bars from our method refer just to the statistical
uncertainties propagated by line of sight kinematics. The sources of
systematic error (anisotropy, external convergence and PSF) are
briefly touched upon case by case and incorporated as a second
uncertainty in the aperture-corrected estimates. We adopt as this
systematic uncertainty $5\%$ of the mean value, even though this might
be too pessimistic.  Finally, we examine the consequences of applying
our single power-law inference to broken power-law density profiles.

\subsection{The Cosmic Horseshoe Lens (SDSS J1148+1930)}

The Cosmic Horseshoe was discovered by \citet{Be07}. It is a massive
early-type lens at a redshift $z_{\rm l} = 0.44$ surrounded by a loose
group. The extended source galaxy at $z_{\rm s} = 2.381$ \citep{Dy08}
is lensed into a nearly complete Einstein ring. Early studies used
estimates of the velocity dispersion derived from low resolution
spectra.

Recently, \citet{Sp11} used high resolution spectra derived from
X-Shooter on the {\it Very Large Telescope} to provide line of sight
kinematic profiles out to the effective radius. Using the Jeans
equations, they find a logarithmic density slope of $\gamma =
1.72^{+0.05}_{-0.06}$. We use their kinematic data ($\sigma = 352\pm19$
inside a circularised aperture of 0.84$^{\prime\prime}$)
with a de Vaucouleurs luminous profile (with an effective radius of
1.96$^{\prime\prime}$; the Einstein radius is 5.1$^{\prime\prime}$),
to get an estimated value of $\gamma=1.76\pm0.08$ without aperture
corrections and $1.80\pm0.04\pm 0.09$ when aperture corrections are included.
This is perfectly compatible with the \citet{Sp11} estimate
within the uncertainties. It is worth remarking that their Jeans analysis
is fit to the line-of-sight velocity dispersion profile, neglecting any
internal rotation. Our method exploits the whole velocity second
moment $\langle\overline{v^{2}}_{\rm los}\rangle_{L}$, (i.e. pressure
and rotation), which is slightly higher than the velocity dispersion
alone. Moreover, we rely just on the surface brightness profile
$\Sigma(R)$ instead of using an approximation for the deprojected
density, which they are forced to do so as to employ the Jeans
equations.

\subsection{The Jackpot Lens (SDSSJ0946+1006)}

The Jackpot Lens was discovered by \citet{Bo08} and subsequently
analysed by \citet{Ga08}. The system is unusual in that it possesses two
Einstein rings, as two different high redshift sources are lensed
by the same foreground elliptical galaxy at a redshift of $z_{\rm l} =
0.22$.

With the photometric, kinematic, and lensing data provided by
\citet{Au09} and \citet{Au10}, our aperture-corrected power-law
exponent is $\gamma=2.01\pm0.15\pm0.1$. This is perfectly consistent with
the estimate $2.01\pm0.18$ of \citet{Au10}, where the uncertainties
from the PSF are fully accounted for. \citet{So12} use a strong prior
on $\gamma$ obtained by more detailed lensing analyses and information
from both sources to find $\gamma = 1.98 \pm 0.03$, in good agreement
with the previous estimates. Using their double Sersic photometry, our
method yields $\gamma=2.06\pm0.04\pm0.1$ if we use the inner Einstein
radius.  On the other hand, their Jeans analysis approximates the
luminous profile with two cored pseudo-Jaffe profiles, but it is not
immediately clear how a flat-top luminous density can be supported
self-consistently in a cusped potential like the scale-free ones, so
the underlying dynamical model may be unphysical~\citep{An09}.

The value $\gamma = 2.20\pm0.03$ estimated by \cite{Ve10} is obtained
by examining a small region around the inner Einstein radius, whereas
our method and the Jeans analysis in \citet{So12} provide an estimate
of the global density exponent on the whole radial range out to the
Einstein radii. This discrepancy hints that the true density profile
is not a simple power-law. One possibility is that the analysis by
\citet{Ve10} probes the density profile where both baryons and dark
matter are important, and the assumption of a single power-law is too
facile.

\subsection{B1608+656}

B1608+656 \citep{mye95,sne95,fas96} consists of an early-type galaxy
lens with a small companion at $z_{\rm l}=0.63.$ The source is an
active galaxy at $z_{\rm s}=1.394.$

In \citet{koo03}, the lensing morphology and time delays of the images
were used to constrain the lensing potential. This was subsequently
refined by incorporating kinematic information in \citet{suy10},
which allowed them to break the degeneracies between density slope and
Hubble constant that would otherwise afflict the problem
\citep{wuc02}. Their complete analysis, assuming Jaffe or Hernquist
luminous densities, gives $\gamma=2.08\pm0.03$. We use the same data as
\citet{suy10} ($\sigma = 260\pm15$ in a 0.42$^{\prime\prime}$ aperture,
the effective radius is 0.6$^{\prime\prime}$, and the Einstein radius of
the main lens is 0.83$^{\prime\prime}$) to obtain $\gamma=2.39\pm0.09$
for the uncorrected estimate and $\gamma=2.14\pm0.06\pm0.11$ including
aperture corrections. As with the Horseshoe, this estimate is consistent
with but slightly larger than previous estimates, although the \citet{suy10}
result is mostly driven by the lensing information from the extended
source structure that is ignored in our analysis.

\begin{figure}
	\centering \includegraphics[width=0.4\textwidth]{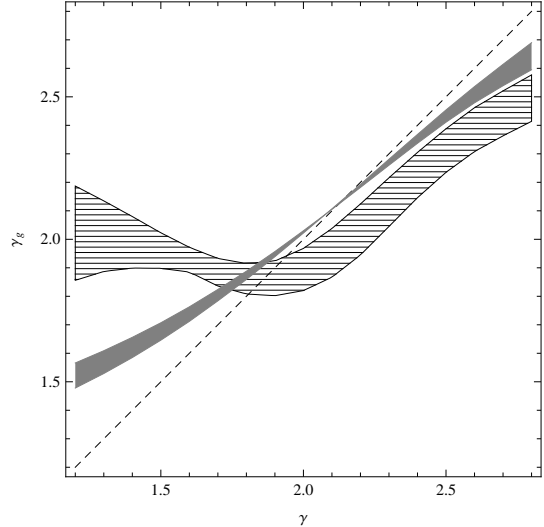}
\caption{\small{The globally-estimated density exponent
    $\gamma_{\rm g}$ versus the inner power-law exponent $\gamma$ of
    the broken power-law model, as given in eq~(\ref{eq:mock}).
    The dashed band illustrates the model prediction when the aperture
    is very large (between 0.25 and 1 times the scale radius of the
    broken power-law), while the grey band is for a small aperture that
    is 5 to 20 times smaller than the effective radius. The widths of
    the bands demonstrate the effects of changing the size of the
    effective radius of the luminous tracer for fixed aperture and
    scale radius; for large effective radii the steep outer halo is
    given more weight and $\gamma_{\rm g}$ is more discrepent from the
    true value $\gamma$.}}
\label{fig:break}
\end{figure}

\subsection{Extensions}

So far, we have assumed a power-law total density. Whether justifiable
or not, this assumption is widely used out to length scales an order of
magnitude greater than the effective radius $r_{\rm e}$ or Einstein
radius $\RE$~\citep[cf][]{gav07,hum10}. However, the normalisation from
lensing is set by the projected mass within the (cylindrical) Einstein
radius, which in turn includes mass at large (spherical polar) radii
where the density profile may fall off differently from any inner
power-law.

As an example, let us consider the broken power-law density
distribution
\begin{equation}
\rho(r)=\frac{\rho_{0}(r/r_{\rm s})^{-\gamma}}{(1+r^{2}/r_{\rm s}^{2})^{(3-\gamma)/2}}\ .
\label{eq:mock}
\end{equation}
This behaves like $\rho \propto r^{-\gamma}$ for $r<r_{\rm s}$, while
$\rho \rightarrow r^{-3}$ for $r\gg r_{\rm s}$.  The appearance of an
explicit scale-radius $r_{\rm s}$ means that the lensing and virial
equations are not enough to fix all the independent parameters. The
virial constraint is then just another piece of information to be
incorporated into the lensing decomposition, albeit now with more
complicated kernels than in eqs ~(\ref{eq:virialens}) and
(\ref{eq:btbeta}).

However, we may ask if approximating this profile with a global
power-law $\rho=\rho_{1}(r/r_{1})^{-\gamma_{\rm g}}$ gives a
reasonable representation of the inner parts of the system. The virial
estimates can help us answer this question. In particular, we find the
global power-law index $\gamma_{\rm g}$ by fixing $\RE$ and
$\langle\overline{v^{2}}_{\rm los}\rangle_{L<R_{\rm ap}}$ to the ones
given by the double-power law profile of eq~(\ref{eq:mock}).

Figure \ref{fig:break} shows bands of solutions corresponding to the
physically plausible range of halo scale lengths with $5 \leq r_{\rm
  s}/\RE \leq 20$.  The dashed curve represents situations when the
aperture is large (here, 5 times the Einstein radius, or 0.25 to 1 times
the scale radius). It is clear that for such large apertures the
power-law approximation for the density profile is inadequate, with the
central slope being grossly over-estimated for small values of $\gamma$
and systematically under-estimated for large values of gamma. This trend
persists when the aperture is small (the grey curve, where the aperture
is equal to the effective radius, i.e., 5 to 20 times smaller than the
scale radius) but with a significantly smaller bias.

The bias of the estimator $\gamma_{\rm g}$ is always towards
isothermality and in the case of small apertures (e.g., consistent
with the apertures used for observations) the bias tends to favour
$\gamma_{\rm g} \approx 2.1$. This is remarkably close to the mean
power-law index inferred for the SLACS lenses \citep{Ko09,Au10,Ba11},
although power-law indices inferred without the use of dynamics yield
similar results \citep[e.g.,][]{Do06}. Nevertheless, the magnitude of 
the bias is also smallest at $\gamma_{\rm g} \approx 2.1$ and suggests
that the power-law index estimates from lensing and dynamics robustly
describe the central density profile even if the true mass density
distribution is a broken power-law.

\section{Conclusions}

We have presented a simple method to combine strong lensing and
stellar dynamics without any need of prescribing the poorly known
velocity dispersion anisotropy. Using measurements of the Einstein
ring radius, the photometric profile and the global kinematics, we
have shown how to obtain the logarithmic slope of the density
$\gamma$.  When the kinematics are measured on a small aperture, then
the density slope $\gamma$ and the velocity anisotropy $\beta$ are
intertwined, but procedures have been developed to characterise the
influence of anisotropy robustly. Above all, our algorithm has the
significant advantage of simplicity -- all that is needed is the
numerical solution of a single non-linear equation to obtain an
estimate for the density slope $\gamma$. We have verified that our
algorithm works well by comparison with analyses in the literature for
three well-studied lenses.

The set of hypotheses used here is common to a majority of Jeans
equation analyses, but the lack of any need for marginalisation over
velocity anisotropy $\beta$ in our method allows us to separate the
statistical uncertainties from systematics. Also, our formulae require
only a profile for the surface brightness, avoiding any additional
uncertainty from deprojection inherent in applications of the Jeans
equations.

Why does our simple method work so well? A major reason is that
velocity anisotropy is never significant at the centres of early-type
galaxies.  If the kinematics are known within a finite aperture
radius, then anisotropy can affect our result. However, this
correction is always small, and isotropic models give a good guide as
to its magnitude. The fact that the virial theorem is independent of
the velocity anisotropy enables us to sidestep difficulties that Jeans
or phase-space modelling must necessarily confront.

The main limitation that this method shares with Jeans analysis is
observational: aperture-averaged line of sight motions must be
suitably dealt with to account for different possible sources of
bias. However, when care is taken in assessing these effects, our
procedure gives the density slope $\gamma$ in terms of directly
observable quantities and can be used as a robust and quick method of
carrying out lensing plus dynamics decompositions.

\vspace{1cm}
\noindent
AA thanks the Science and Technology Facility Council and the Isaac
Newton Trust for the award of a studentship. We wish to thank Nicola
C. Amorisco, Iulia T. Simion and Vasily Belokurov for significant
feedback on the manuscript, as well as Alessandro Sonnenfeld for
discussing some details of his analysis with us. The anonymous referee
gave useful comments that helped us improve the paper.

\label{lastpage}

\begin{thebibliography}{}
\bibitem[Agnello \& Evans (2012)]{Ag12} Agnello, A. \& Evans, N.W. 2012,
  \apjl, 754, L39 

\bibitem[An \& Evans(2009)]{An09} An, J.~H., \& Evans, N.W.\ 2009,
  \apj, 701, 1500

\bibitem[Auger et al.(2009)]{Au09} Auger, M.~W., Treu, T., 
Bolton, A.~S., et al.\ 2009, \apj, 705, 1099 

\bibitem[Auger et al.(2010)]{Au10} Auger, M.~W., Treu, T., Bolton,
  A.~S., et al.\ 2010, \apj, 724, 511

\bibitem[Barnab{\`e} \& Koopmans(2007)]{Ba07} Barnab{\`e}, M., \&
  Koopmans, L.~V.~E.\ 2007, \apj, 666, 726

\bibitem[Barnab{\`e} et al.(2011)]{Ba11} Barnab{\`e}, M., 
Czoske, O., Koopmans, L.~V.~E., Treu, T., \& Bolton, A.~S.\ 2011,
\mnras, 415, 2215 

\bibitem[Belokurov et al.(2007)]{Be07} Belokurov, V., Evans, 
N.~W., Moiseev, A., et al.\ 2007, \apjl, 671, L9 

\bibitem[Bolton et al.(2008)]{Bo08} Bolton, A.~S., Burles, S.,
  Koopmans, L.~V.~E., et al.\ 2008, \apj, 682, 964

\bibitem[Ciotti \& Morganti(2010)]{Ci10} Ciotti, L., \& Morganti,
  L.\ 2010, \mnras, 408, 1070

\bibitem[Dobke \& King(2006)]{Do06} Dobke, B.~M., \& King, L.~J.\ 2006, \aap, 460, 647 

\bibitem[Dye et al.(2008)]{Dy08} Dye, S., Evans, N.~W., Belokurov,
  V., Warren, S.~J., \& Hewett, P.\ 2008, \mnras, 388, 384

\bibitem[Falco et al.(1985)]{fal85} Falco, E.~E., Gorenstein, 
M.~V., \& Shapiro, I.~I.\ 1985, \apjl, 289, L1 

\bibitem[Fassnacht et al.(1996)]{fas96} Fassnacht, C.~D., 
Womble, D.~S., Neugebauer, G., et al.\ 1996, \apjl, 460, L103 

\bibitem[Gavazzi et al.(2007)]{gav07} Gavazzi, R., Treu, T., 
Rhodes, J.~D., et al.\ 2007, \apj, 667, 176 

\bibitem[Gavazzi et al.(2008)]{Ga08} Gavazzi, R., Treu, T., 
Koopmans, L.~V.~E., et al.\ 2008, \apj, 677, 1046 

\bibitem[Gerhard et al.(2001)]{Ge01} Gerhard, O., Kronawitter, A.,
  Saglia, R.~P., \& Bender, R.\ 2001, \aj, 121, 1936

\bibitem[Humphrey 
\& Buote(2010)]{hum10} Humphrey, P.~J., \& Buote, D.~A.\ 2010, \mnras, 403, 2143 
\bibitem[Koopmans et al.(2003)]{koo03} Koopmans, L.~V.~E., 
Treu, T., Fassnacht, C.~D., Blandford, R.~D., 
\& Surpi, G.\ 2003, \apj, 599, 70 

\bibitem[Koopmans et al.(2009)]{Ko09} Koopmans, L.~V.~E., Bolton, A.,
  Treu, T., et al.\ 2009, \apjl, 703, L51

\bibitem[Myers et al.(1995)]{mye95} Myers, S.~T., Fassnacht, 
C.~D., Djorgovski, S.~G., et al.\ 1995, \apjl, 447, L5 

%\bibitem[Press et al.(1986)]{Pr07} Press, W.~H., Flannery, B.~P., \& Teukolsky, S.~A.\ 1986, Cambridge: University Press, 1986,  

\bibitem[Saglia et al.(1993)]{Sa93} Saglia, R.~P., Bertin, G.,
  Bertola, F., et al.\ 1993, \apj, 403, 567

\bibitem[Schneider, Ehlers \& Falco(1992)]{Sc92} Schneider P., Ehlers J.,
  Falco E.E., 1992, Gravitational Lenses, Springer Verlag, New York

\bibitem[Snellen et al.(1995)]{sne95} Snellen, I.~A.~G., de 
Bruyn, A.~G., Schilizzi, R.~T., Miley, G.~K., 
\& Myers, S.~T.\ 1995, \apjl, 447, L9 

\bibitem[Sonnenfeld et al.(2012)]{So12} Sonnenfeld, A., 
Treu, T., Gavazzi, R., et al.\ 2012, \apj, 752, 163 

\bibitem[Spiniello et al.(2011)]{Sp11} Spiniello, C., Koopmans,
  L.~V.~E., Trager, S.~C., Czoske, O., \& Treu, T.\ 2011, \mnras, 417,
  3000

\bibitem[Suyu et al (2010)]{suy10} Suyu, S. H., et al. 2010 \apj, 711,
  201

\bibitem[Treu \& Koopmans(2004)]{Tr04} Treu, T., \& Koopmans,
  L.~V.~E.\ 2004, \apj, 611, 739

\bibitem[Vegetti et al.(2010)]{Ve10} Vegetti, S., Koopmans, L.~V.~E.,
  Bolton, A., Treu, T., \& Gavazzi, R.\ 2010, \mnras, 408, 1969

\bibitem[Wucknitz(2002)]{wuc02} Wucknitz, O.\ 2002, \mnras, 
332, 951 

\end{thebibliography}
\end{document}